\documentstyle[aps,prb,epsfig]{revtex} 
\begin{document} 
\draft 
\twocolumn 
 
\title 
{Quantum Conductors in a Plane} 
\author{Philip Phillips$^1$, Subir Sachdev$^2$, Sergey Kravchenko$^3$, and Ali Yazdani$^1$} 
 
%
\address 
{$^1$Loomis Laboratory of Physics,University of Illinois at 
Urbana-Champaign, 1100 W.Green St., Urbana, IL, 61801-3080, 
$^2$Department of Physics, P. O. Box 208120, Yale University, New 
Haven, CT. 06520, $^3$Department of Physics, Northeastern 
University, Boston, MA. 02115} 
 
%
\maketitle

\columnseprule 0pt \narrowtext 
 
When electrons are confined to move in a plane, strange things 
happen. For example, under normal circumstances, they are not 
expected  to conduct electricity at low temperatures. The absence 
of electrical conduction in two dimensions at zero temperature has 
been one of the most cherished paradigms in solid state 
physics\cite{abrahams}.  In fact, the 1977 physics Nobel prize was 
awarded, in part, for the formulation of the basic principle on 
which this result is based. However, recent 
experiments\cite{kravchenko} on a dilute electron gas confined to 
move at the interface between two semiconductors pose a distinct 
counterexample to the standard view.  Transport measurements 
reveal\cite{kravchenko} that as the temperature is 
lowered, the resistivity drops without any signature of the 
anticipated up-turn as required by the standard account.  It is 
the possible existence of a new conducting state and hence
a new quantum phase transition in two dimensions 
that is the primary focus of this session. In the absence of a 
magnetic field, the only quantum phase transition known to exist 
in two dimensions (2D) that involves a conducting phase is the 
insulator-superconductor transition\cite{goldman}. Consequently, 
this session focuses on the general properties of quantum phase 
transitions, the evidence for the new conducting state in a 2D 
electron gas and the range of phenomena that can occur in 
insulator-superconductor transitions. 
 
Unlike classical phase transitions, such as the melting of ice, 
all quantum phase transitions occur at the absolute zero of 
temperature.  While initially surprising, this state of affairs is 
expected as quantum mechanics is explicitly a zero-temperature 
theory of matter.  As such, quantum phase transitions are not 
controlled by changing system parameters such as the temperature 
as in the melting of ice, but rather by changing some external 
parameter such as the number of defects or the magnitude of an 
applied magnetic field. 
In all instances, the underlying quantum mechanical states are 
transformed between ones that either look different topologically 
or have distinctly different magnetic properties. Two examples of 
quantum phase transitions are the disorder-induced metal-insulator 
transition and the insulator-superconductor transition.  
In a clean 
crystal, electrons form perfect Bloch waves or traveling waves 
and move unimpeded throughout the crystal.  When defects (disorder) are present, 
electrons can become characterized by exponentially-decaying 
states which cannot carry current at 
zero-temperature because of their confined spatial extent. In a 
plane, the localization principle\cite{abrahams} establishes that 
as long as electrons act independently, only localized states form 
whenever disorder is present. However, if for some strange reason, 
such as they are attracted through a third party to one another, 
electrons can form pairs.
Such pairs constitute the charge 
carriers in a superconductor and are called Cooper pairs. 
Superconductors are perfect conductors of electricity and therfore 
have a vanishing resistance.  However, formation of Cooper pairs is not a sufficient 
condition for superconductivity. If one envisions dividing a 
material into partitions, 
insulating behaviour obtains if each partition at each snapshot in 
time has the same number of Cooper pairs.  That is, the state is 
static. However, if the number of pairs fluctuates between 
partitions, transport of Cooper pairs is possible and 
superconductivity obtains.

The fundamental physical principle that drives all quantum phase 
transitions is quantum uncertainty or quantum entanglement.  A 
superconductor can be viewed as an entangled state containing all 
possible configurations of the Cooper pairs. Scattering a single 
Cooper pair would require disrupting each configuration in which 
that Cooper pair resides. Since each Cooper pair exists in each 
configuration (of which there are an infinite number), such a 
scattering event is highly improbable. We refer to a 
superconducting state then as possessing phase coherence, that is 
rigidity to scattering. Insulators lack phase coherence. 
In the insulating state,
the certainty that 
results in the particle number within each partition is 
counterbalanced by the complete loss of phase coherence.  
In a superconductor, phase certainty gives rise 
to infinite uncertainty in the particle number. Consequently, the 
product of the number uncertainty times the uncertainty in phase 
is the same on either side of the transition as dictated by the 
Heisenberg uncertainty principle. 
In essence, quantum uncertainty 
is to quantum phase transitions what thermal agitation is to 
classical phase transitions.  Both transform matter from one state 
to another. 	
  
\begin{figure}
\begin{center}
\epsfig{file=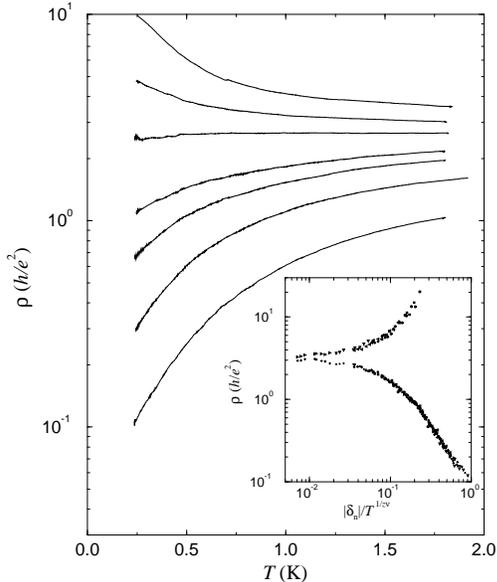,width=6.5cm}
\caption{Resistivity ($\rho$) vs.~temperature for two-dimensional 
electrons in silicon in zero magnetic field and at different 
electron densities (n) (from top to bottom:  0.86, 0.88, 0.90, 
0.93, 0.95, 0.99, and $1.10\times10^{11}$ per cm$^2$.  Collapse of 
the data onto two distinct scaling curves above and below the 
critical transition density ($n_c$) is shown in the inset.  Here 
$\delta=(n-n_c)/n_c$, $z=0.8$ and $\nu=1.5$.} 
\label{fig2}
\end{center}
\end{figure}

In the experiments revealing the new 
conducting phase, the tuning parameter is the concentration of 
charge carriers\cite{kravchenko,popovic,pepper,shahar}.  For 
negatively-charged carriers, such as electrons, a positive bias 
voltage is required to adjust the electron 
density\cite{kravchenko,popovic}--the more positive, the higher 
the electron density.  Subsequently, if the electrons are confined 
to move laterally at the ultra-thin (25\AA) interface between two 
semiconductors, transport will be two-dimensional as it is 
confined to a plane. Devices of this sort constitute a special 
kind of transistor, not too dis-similar from those used in desktop 
computers. As illustrated in Fig.~(\ref{fig2}), when the electron 
density is slowly increased beyond $\approx 10^{11}/cm^2$, the 
resistivity changes from increasing (insulating behavior) to 
decreasing as the temperature decreases, the signature of 
conducting behavior. 
At the transition between these two limits, 
the resistivity is virtually independent of temperature.  While it 
is still unclear ultimately what value the resistivity will 
acquire at zero temperature, the marked decrease in the 
resistivity above a certain density is totally unexpected and 
more importantly not predicted by any theory. Whether we can 
correctly conclude that a zero-temperature transition exists 
between two distinct phases of matter is still not settled, 
however.  Nonetheless, the data do possess a feature common to 
quantum phase transitions\cite{fisher}, namely scale 
invariance.  In this context, scale invariance simply implies that 
the data above the flat region in Fig.~(\ref{fig2}) all look 
alike. This also holds for the data below the flat region in 
Fig.~(\ref{fig2}). As a consequence, the upper and lower family of 
resistivity curves at various densities can all be made to 
collapse onto just two distinct curves by scaling each curve with 
the same density-dependent scale factor. The resultant curves have 
slopes of opposite sign as shown in the inset of Fig. 
(\ref{fig2}). It is difficult to reconcile this bi-partite 
structure unless the two phases are in fact distinct electrically 
at zero temperature. 
 
These experiments lead naturally to the question, what is so 
special about the density regime probed. We know definitively that 
at high and ultra-low densities, a 2D electron gas is localized by 
disorder.  Because the Coulomb interaction decays as $1/r$ 
(with $r$ the separation between the electrons) whereas the 
kinetic energy decays as $1/r^2$, Coulomb interactions dominate at 
low density.  At sufficiently low electron densities, the electrons
form a crystal.  It is precisely between the ultra-low crystalline
limit and 
the non-interacting regime that the possibly new conducting phase 
resides.  This density regime represents one of the 
yet-unconquered frontiers in solid state physics. 
Experimentally, it is clear that whatever happens in this 
intermediate density regime is far from ordinary as evidenced by 
the observed destruction\cite{krav1} of the conducting phase by an 
applied in-plane magnetic field.  As an in-plane magnetic field 
can only polarize the spins, the conducting phase is highly 
sensitive to the spin state, a key characteristic of 
superconductivity. 
 
Experimentally, a direct transition from a superconductor to an 
insulator in 2D has been observed by two distinct mechanisms.  The 
first is simply by decreasing the thickness of the 
sample\cite{goldman}. This effectively changes the scattering 
length and hence is equivalent to changing the amount of disorder. 
As a result, Cooper pairs remain intact throughout the transition. 
While single electrons are localized by disorder, Cooper pairs in 
a superconducting state are not.  Under normal circumstances,
Cooper pairs give rise to a zero resistance state at
$T=0$.  The 
second means by which a superconducting state can be transformed 
to an insulator in 2D is by applying a perpendicular magnetic 
field\cite{yazdani,fisher}. A perpendicular magnetic 
field creates resistive excitations called vortices (the dual of 
Cooper pairs) which frustrate the onset of global phase coherence. 
Surprisingly, howeer, in both the disorder\cite{jaeger} and magnetic field-tuned 
transitions\cite{yazdani,mason}, the
resistivity has been observed to flatten on the 
`superconducting' side. 
The non-vanishing of the resistivity is indicative of a lack of phase 
coherence. Phase fluctuations are particularly strong in 2D and 
are well-known to widen the temperature regime over which the 
resisitivity drops to zero.  
However, the precise origin of the flattening of the resistivity (an
indication of a possible metallic state) at low temperatures is not
known. 
 
Ultimately, the resolution of the experimental puzzles raised here 
must be settled by further experiments. But a natural question 
that arises is, are the two phenomena related? This question is 
particularly germane\cite{phillips} because the only excitations 
proven to survive the localizing effect of disorder in 2D are 
Cooper pairs. It is for partly this simple reason\cite{phillips} 
and other more complex arguments\cite{zhang,bk} that 
superconductivity has been proposed to explain the new conducting 
state in 2D. Because phase fluctuations create a myriad of options 
(`metal' or superconductor at T=0) for Cooper pairs in a plane, 
measurements sensitive to pair formation must augment the standard 
transport measurements to definitively settle whether Cooper pair 
formation is responsible for new conducting state in a 2D 
electron gas. But maybe some yet-undiscovered\cite{chak} 
conducting spin singlet state exists that can survive the 
localizing effect of disorder. But maybe not and possibly only 
`classical' trapping effects\cite{maslov} are responsible for the decrease of 
the resistivity on the conducting side. While the former cannot be 
ruled out, the latter seems unlikely as new 
experiments\cite{krav2} reveal the new conducting phase is tied
to the formation of a Fermi surface and related to
the plateau transitions\cite{hanein99} in the quantum Hall effect.
This implies that indeed a 
deep quantum mechanical principle is responsible for the new 
conducting state, perhaps as has been suggested\cite{phillips} 
that the proximity of the new conducting phase to a 
strongly-correlated insulator mediates pairing as in copper-oxide superconductors.

\end{document}